%% file: main.tex
\documentclass[sigconf]{acmart}
\usepackage{booktabs} 
\usepackage{spreadtab}
\usepackage{amsmath}
\usepackage{float}
\usepackage{graphicx}
\usepackage{caption}
\usepackage{color}
\usepackage{subcaption}
\DeclareCaptionSubType*[Alph]{table}
\DeclareCaptionLabelFormat{mystyle}{Table~\bothIfFirst{#1}{ ̃}#2}
\usepackage{multirow}

\copyrightyear{2019} 
\acmYear{2019} 
\setcopyright{acmcopyright}
\acmConference[SIGIR '19]{Proceedings of the 42nd International ACM SIGIR Conference on Research and Development in Information Retrieval}{July 21--25, 2019}{Paris, France}
\acmBooktitle{Proceedings of the 42nd International ACM SIGIR Conference on Research and Development in Information Retrieval (SIGIR '19), July 21--25, 2019, Paris, France}
\acmPrice{15.00}
\acmDOI{10.1145/3331184.3331303}
\acmISBN{978-1-4503-6172-9/19/07}

\begin{document}

\title{Deeper Text Understanding for IR with Contextual Neural Language Modeling}

\author{Zhuyun Dai}
\affiliation{%
   \institution{Carnegie Mellon University}
   }
\email{zhuyund@cs.cmu.edu}

\author{Jamie Callan}
\affiliation{%
   \institution{Carnegie Mellon University}
   }
\email{callan@cs.cmu.edu}

\begin{abstract}

\end{abstract}




\begin{abstract}
Neural networks provide new possibilities to automatically learn complex language patterns and query-document relations. Neural IR models have achieved promising results in learning query-document relevance patterns, but few explorations have been done on understanding the text content of a query or a document. This paper studies leveraging a recently-proposed contextual neural language model, BERT, to provide deeper text understanding for IR. Experimental results demonstrate that the contextual text representations from BERT are more effective than traditional word embeddings. Compared to bag-of-words retrieval models, the contextual language model can better leverage language structures, bringing large improvements on queries written in natural languages. Combining the text understanding ability with search knowledge leads to an enhanced pre-trained BERT model that can benefit related search tasks where training data are limited.
\end{abstract}


\maketitle

\input{sections/introduction}
\input{sections/related_work}

\input{sections/model.tex}
\input{sections/experimental_setup.tex}

\input{sections/evaluation.tex}

\input{sections/conclusion.tex}

\section{ACKNOWLEDGMENTS}
This research was supported by NSF grant IIS-1815528. Any opinions, findings, and conclusions in this paper are the authors' and do not necessarily reflect those of the sponsors.

\bibliographystyle{ACM-Reference-Format}
\bibliography{bibliography}

\end{document}

%% file: sections/introduction.tex
\section{Introduction}

Text retrieval requires understanding document meanings and the search task.
Neural networks are an attractive solution because they can acquire that understanding from raw document text and training data. Most neural IR methods focus on learning query-document relevance patterns, that is, knowledge about the search task. However, learning only relevance patterns requires large amounts of training data, and yet still doesn't generalize well to tail queries~\cite{K-NRM} or new search domains~\cite{dai2018convolutional}.
These issues make pre-trained, general-purpose text understanding models desirable.


Pre-trained word representations such as word2vec~\cite{word2vec} have been widely used in neural IR. They are learned from word co-occurrence in a large corpus, providing hints about synonyms and related words. But word co-occurrence is only a shallow bag-of-words understanding of the text.  
Recently, we have seen rapid progress in text understanding with the introduction of pre-trained neural language models such as ELMo~\cite{Peters:2018} and BERT~\cite{devlin2018bert}. 
Different from traditional word embeddings, they are \emph{contextual}  -- the representation of a word is a function of the entire input text, with word dependencies and sentence structures taken into consideration. The models are \emph{pre-trained} on a large number of documents so that the contextual representations encode general language patterns. Contextual neural language models have outperformed traditional word embeddings on a variety of NLP tasks~\cite{Peters:2018, devlin2018bert}.

The deeper text understanding of contextual neural language models brings new possibilities to IR. This paper explores leveraging BERT (Bidirectional Encoder Representations from Transformers)~\cite{devlin2018bert} for ad-hoc document retrieval. BERT is a state-of-the-art neural language model. It also fits well with search tasks. 
BERT is trained to predict the relationship between two pieces of text (typically sentences); and its attention-based architecture models the local interactions of words in text$_1$ with words in text$_2$.  It can be viewed as an interaction-based neural ranking model~\cite{DRMM}, thus minimal search-specific architectural engineering is required.  
 




This paper explores the effect of BERT's language understanding on ad-hoc document retrieval. It examines BERT models on two ad-hoc retrieval datasets with different characteristics.  Experiments show that fine-tuning pre-trained BERT models with a limited amount of search data can achieve better performance than strong baselines. In contrast to observations from traditional retrieval models, longer natural language queries are able to outperform short keywords queries by large margins with BERT. Further analysis reveals that stopwords and punctuation, which are often ignored by traditional IR approaches, play a key role in understanding natural language queries by defining grammar structures and word dependencies. Finally, enhancing BERT with search knowledge from a large search log produces a pre-trained model equipped with knowledge about both text understanding and the search task, which benefits a related search task where labeled data are limited.


%% file: sections/related_work.tex
\section{Related Work}\label{section:related-work}


Recent neural IR models have made promising progress in learning query-document relevance patterns. One line of research learns text presentations tailored for the search task~\cite{K-NRM, dai2018convolutional, WeakNeuIR} with search signals from click logs~\cite{K-NRM, dai2018convolutional} or pseudo-relevance feedback~\cite{WeakNeuIR}. Another line of research designs neural architectures to capture diverse matching features such as exact match signals~\cite{DRMM} and passage-level signals~\cite{pang2017deeprank}. How to understand the text content of the query/document is less explored. Most neural IR models represent text with word embeddings such as Word2Vec~\cite{word2vec}. 

Contextual neural language models are proposed to improve traditional word embeddings by incorporating the context~\cite{Peters:2018, devlin2018bert}. 
One of the best performing neural language models is BERT~\cite{devlin2018bert}.  
BERT is pre-trained on large-scale, open-domain documents to learn general patterns in a language. Pre-training tasks include predicting words within a sentence and the relationship of two sentences. BERT has advanced the state-of-the-art on a variety of NLP tasks, including passage ranking tasks~\cite{nogueira2019passage}. Its effectiveness on standard document retrieval tasks remains to be studied.


%% file: sections/model.tex
\section{Document Search with BERT}\label{sec:model}

\begin{figure}[!t]
\centering
\includegraphics[width=0.75\linewidth]{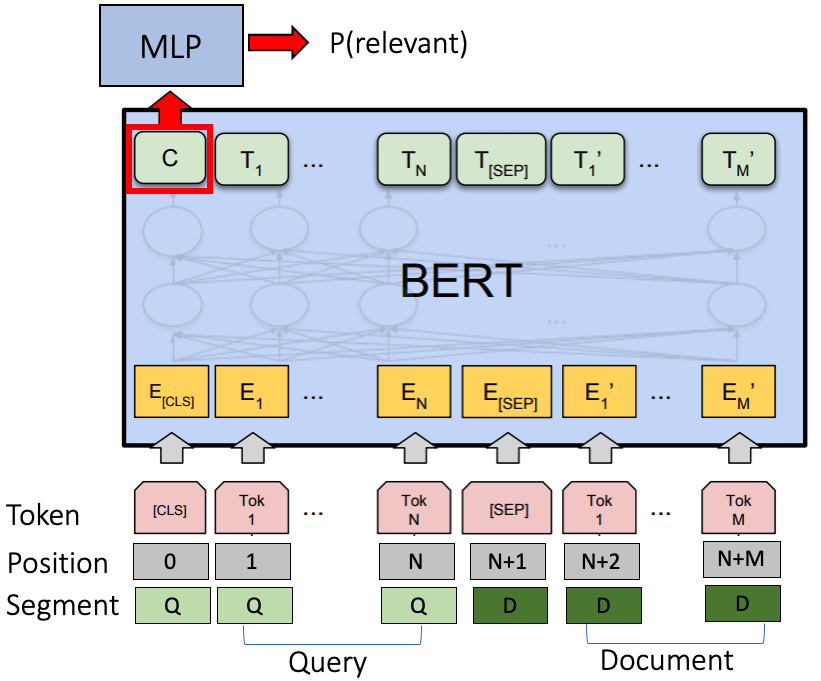}
\caption{BERT sentence pair classification architecture~\cite{devlin2018bert}. 
}
\label{fig:bert}
\end{figure}

This work uses an off-the-shelf BERT architecture, the sentence pair classification architecture described by~\citet{devlin2018bert}, as shown in Figure~\ref{fig:bert}. The model takes as input the concatenation of the query tokens and the document tokens, with a special token `[SEP]'  separating the two segments. Tokens are embedded into embeddings. To further separate the query from the document, segment embeddings `Q' (for query tokens) and `D' (for document tokens) are added to the token embeddings. To capture word order, position embeddings are added. The tokens go through several layers of \emph{transformers}. At each layer, a new contextualized embedding is generated for each token by weighted-summing all other tokens' embeddings. The weights are decided by several attention matrices (multi-head attention). Words with stronger attentions are considered more related to the target word. Different attention matrices capture different types of word relations, such as exact match and synonyms . The attention spans over the query and the document so that query-document interactions are taken into consideration. Finally, the output embedding of the first token is used as a representation for the entire query-document pair. It is fed into a multi-layer perceptron (MLP) to predict the possibility of relevance (binary classification).
The model is initialized with a pre-trained BERT model\footnote{We use uncased BERT-Base model from \url{https://github.com/google-research/bert}} to leverage the pre-trained language model, while the last MLP layer is learned from scratch. During training, the entire model is tuned to learn more IR-specific representations.

\textbf{Passage-Level Evidence.} Applying BERT to long documents causes increasing memory usage and run time due to the complexity in interacting every pair of tokens. The sentence-trained model may also be less effective on long text. We adopt a simple passage-level approach for document retrieval. We split a document into overlapping passages. The neural ranker predicts the relevance of each passage independently.  document score is the score of the first passage (\texttt{BERT-FirstP}), the best passage (\texttt{BERT-MaxP}), or the sum of all passage scores (\texttt{BERT-SumP}). For training, passage-level labels are not available in this work. We consider all passages from a relevant document as relevant and vice versa. When the document title is available, the title is added to the beginning of every passage to provide context.

\textbf{Augmenting BERT with Search Knowledge.}  
Some search tasks require both general text understanding (e.g. Honda is a motor company) and more-specific search knowledge (e.g. people want to see special offers about Honda). While pre-trained BERT encodes general language patterns, the search knowledge must be learned from labeled search data.  Such data are often expensive to obtain and take time to collect. It is desirable to have a pre-trained ranking model that is equipped with both language understanding knowledge and search knowledge. We augment BERT with search knowledge by tuning it on a large search log. The rich search knowledge is expected to benefit related search tasks where only dozens or hundreds of labeled search examples are available. 


\textbf{Discussion.} 
Only small adaptations are made to apply BERT to search tasks:  
a passage-based approach to handle long documents, and a concatenation approach to handle multiple document fields.  Our goal is to study the value of BERT's contextualized language models for search, not to make major 
extensions to the architecture.

%% file: sections/experimental_setup.tex
\section{Experimental Setup} \label{section:exp-setup}

\begin{table}
\caption{Example of Robust04 search topic (Topic 697).}\label{tab:query-example}
\small
\vspace{-0.15in}

\begin{tabular}{p{0.15\columnwidth} p{0.76\columnwidth}}
\hline \hline
Title       & air traffic controller                                                                                                                                                                                                                                                  \\
Description & What are working conditions and pay for U.S. air traffic controllers?                                                                                                                                                                                             \\
Narrative   &  Relevant documents tell something about working conditions or pay for American controllers.  Documents about foreign controllers or individuals are not relevant.
\\ \hline \hline
\end{tabular}
\end{table}

\textbf{Datasets.} We use two standard text retrieval collections with different characteristics. \textbf{Robust04} is a news corpus with 0.5M documents and 249 queries. Two versions of queries are included: a short keyword query (\textit{title}) and a longer natural language query (\textit{description}). A \textit{narrative} is also included as the guidance for relevance assessment. An example is shown in Table \ref{tab:query-example}. \textbf{ClueWeb09-B} contains 50M web pages and 200 queries with title and description. Passages are generated using a $150$-word sliding window with a stride of $75$ words. For ClueWeb09-B, document titles are added to the beginning of each passage. 
For augmenting BERT with search data, we follow the domain adaptation setting from ~\citet{dai2018convolutional} and use the same Bing search log sample. The sample contains 0.1M queries and 5M query-document pairs.

\textbf{Baselines and Implementations}. Unsupervised baselines use Indri's bag of words (\texttt{BOW}) and sequential dependency model queries (\texttt{SDM}). 
Learning-to-rank baselines include \texttt{RankSVM} and \texttt{Coor-Ascent} with bag-of-words features~\cite{dai2018convolutional}. Neural baselines include \texttt{DRMM}~\cite{DRMM} and \texttt{Conv-KNRM}. \texttt{DRMM} uses word2vec~\cite{word2vec} to model word soft-match; it was shown to be among the best performing neural models on our two datasets~\cite{DRMM}. \texttt{Conv-KNRM} learns n-gram embeddings for the search task and show strong performance when 
trained on large search logs~\cite{dai2018convolutional}.  
The Bing-adapted \texttt{Conv-KNRM} was the state-of-the-art neural IR model when trained with domain adaptation~\cite{dai2018convolutional}, and is compared to the Bing-augmented BERT. BERT models are based on the implementation released by Google\footnote{https://github.com/google-research/bert}. Baselines use standard stopword removal and stemming; BERT uses raw text. Supervised models are used to re-rank the top 100 documents retrieved by \texttt{BOW} with 5-fold cross-validation. Due to space limit, we only report \textbf{nDCG@20}; similar trends were observed with nDCG@10 and MAP@100. Source code and related resources are released~\footnote{\url{https://github.com/AdeDZY/SIGIR19-BERT-IR}}.

%% file: sections/evaluation.tex
\section{Results and Discussion}

This section studies the effectiveness of BERT on document retrieval tasks, the difference among different types of queries, and the impact of enhancing BERT with search logs.

\input{sections/exp1-overall.tex}

\input{sections/exp2-query.tex}

\input{sections/exp3-enhance.tex}

%% file: sections/exp1-overall.tex
\subsection{Pre-trained BERT for Document Retrieval}

\begin{table}[t]
\caption{Search accuracy on \textbf{Robust04} and \textbf{ClueWeb09-B}. $\dagger$ indicates statistically significant improvements over \texttt{Coor-Ascent} by permutation test with p< 0.05.}\label{tab:overall}

\def\arraystretch{0.9}
\vspace{-0.15in}
\begin{tabular}{l|cc||cc}
\hline \hline
  &  \multicolumn{4}{c}{nDCG@20}\\
\multirow{3}{*}{Model} & \multicolumn{2}{c||}{Robust04} & \multicolumn{2}{c}{ClueWeb09-B} \\
 & \multicolumn{1}{c}{Title  } & \multicolumn{1}{c||}{Description  } & \multicolumn{1}{c}{Title  } & \multicolumn{1}{c}{Description }  \\
                            \hline
\texttt{BOW}                         & 0.417    & 0.409   
 & 0.268  & 0.234  \\
\texttt{SDM}                         & 0.427           & 0.427         & 0.279 & 0.235  \\ \hline
\texttt{RankSVM}                    & 0.420         &   0.435        & 0.289  & 0.245  \\
\texttt{Coor-Ascent}                & 0.427            &     0.441               & \textbf{0.295}  & 0.251  \\ \hline
\texttt{DRMM}                       & 0.422          & 0.412            & 0.275  & 0.245  \\ 
\texttt{Conv-KNRM}                       &  0.416        &  0.406          & 0.270  &  0.242 \\ \hline\hline
\texttt{BERT-FirstP}           & 0.444$^\dagger$            & 0.491$^\dagger$             & 0.286  & \textbf{0.272}$^\dagger$    \\
\texttt{BERT-MaxP}            & \textbf{0.469}$^\dagger$           & \textbf{0.529}$^\dagger$          & 0.293  & 0.262$^\dagger$   \\ 
\texttt{BERT-SumP}        &  0.467$^\dagger$     &   0.524$^\dagger$    &   0.289 &   0.261      \\ 
\hline \hline

\end{tabular}
\end{table}

The ranking accuracy of each ranking method is shown in Table~\ref{tab:overall}.  On Robust04,  BERT models consistently achieve better search accuracy than the baselines, with a $10\%$ margin on query titles and a $20\%$ margin on description queries. On ClueWeb09-B,  BERT is comparable to 
\texttt{Coor-Ascent} on title queries, and better on description queries. The results demonstrate the effectiveness of BERT for document retrieval, especially on description queries. 

Among the neural rankers, \texttt{Conv-KNRM} has the lowest accuracy. \texttt{Conv-KNRM} needs to learn n-gram embeddings from scratch. It is strong when 
trained on a large search log~\cite{dai2018convolutional}, but it tends to overfit when 
trained with only a small amount of data. BERT is pre-trained and is less prone to overfitting. \texttt{DRMM} represents words with pre-trained word embeddings. The better performance of BERT models demonstrates that the contextualized text representations are more effective for IR than bag-of-words embeddings. 

\begin{figure}[tb]
     \centering
     
     \includegraphics[width=0.45\textwidth]{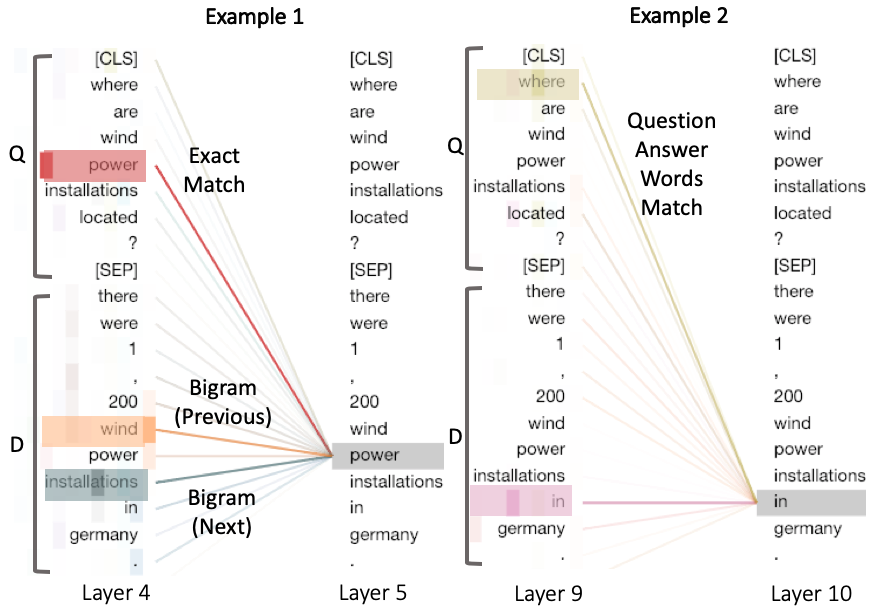}
     \caption{Visualization of BERT. Colors represent different attention heads; deeper color indicates higher attention. } \label{fig:bertviz}
 \end{figure}

\textbf{Sources of effectiveness}. Figure~\ref{fig:bertviz} visualizes two layers from the \texttt{BERT-MaxP} model when predicting the relevance between a description query \textit{`Where are wind power installations located?'} and a sentence \textit{`There were 1,200 wind power installations in Germany'}. Example 1 shows the attention received by the document word \textit{`power'}. The strongest attention comes from \textit{`power'} in the query (query-document exact term matching), and the previous and next word of \textit{`power'} (bi-gram modeling). 
Local matching of words and n-grams have proven to be strong neural IR features ~\cite{DRMM, dai2018convolutional}; BERT is also able to capture them.
Example 2 shows that the document word \textit{`in'} receives the strongest attention from the query word \textit{`where'}. The word \textit{`in'} appears in the context of \textit{`in Germany'}, so it satisfies the \textit{`where'} question. Words like \textit{`in'} and \textit{`where'} are often ignored by traditional IR methods due to their high document frequency in the corpus. This example shows that with a deeper text understanding, these stop words actually provide important evidence about relevance. 
In summary, the advantages of BERT lie in the architecture and the data. The transformer architecture allows BERT to extract a variety of effective matching features. The transformers are already pre-trained on a large corpus, so search tasks with few training data can also benefit from this deep network.

\textbf{Title queries vs. description queries}. \texttt{BERT} models have larger gains on description queries. On Robust04, using description queries with \texttt{BERT-MaxP} brings a $23\%$ improvement over the best title query baseline (\texttt{SDM}). Most other ranking methods only get similar or worse performance on descriptions compared to titles. To the best of our knowledge, this is the first time we see that description queries outperform title queries with such a large margin. On ClueWeb09-B,  \texttt{BERT} manages to shrink the gap between titles and descriptions. 
Although intuitively, description queries should carry richer information, it is hard to fully utilize them in traditional bag-of-words methods due to difficulties in estimating term importance. Our results show that longer natural language queries are indeed more expressive than keywords, and the richer information can be effectively leveraged to improve search using a deep, contextualized neural language model. Further analysis of BERT's ability to understand different types of search queries is given in Section 5.2.

\textbf{Robust04 vs. ClueWeb09-B.} BERT models perform better on Robust04 than on ClueWeb09-B. This is probably due to that Robust04 is closer to the pre-trained model. Robust04 has well-written articles; its queries look for facts that depend largely on understanding text meaning. ClueWeb09-B documents are webpages that include tables, navigation bars, and other discontinuous text. The task also involves web-search specific issues such as page authority. More training data may be required to learn such search-specific knowledge. We investigate this possibility in Section 5.3.

%% file: sections/exp2-query.tex
\subsection{Understanding Natural Language Queries}


This section examines BERT on 3 types of queries that require different levels of text understanding: title, description, and narrative.  To test the effect of grammar structures, a keyword version of description and narrative is generated by removing stopwords and punctuation. To test how BERT understands the logic in narratives, a `positive' version of narrative is generated by removing negative conditions (e.g. \textit{"Not relevant are documents..."}).
Table~\ref{tab:query-types} shows the performance of \texttt{SDM}, \texttt{Coor-Ascent} and \texttt{BERT-MaxP} on Robust04. Supervised methods use narratives to re-rank title query initial results due to low recall of \texttt{BOW} on narratives, which gives narratives 
an advantage over the other types of queries.

\texttt{SDM} works best with titles. \texttt{Coor-Ascent} is moderately better with descriptions and narratives. The two methods weight words solely based on term frequencies, but word importance should depend on the meaning of the entire query. In contrast, \texttt{BERT-MaxP} makes large improvement on longer queries by modeling word meaning and context. Keywords versions perform better than the original query for \texttt{SDM} and \texttt{Coor-Ascent}, because stopwords are noisy to traditional match signals like TF.
In contrast, BERT is more effective on the original natural languages queries. Although stopwords and punctuation do not define the information need, they build the structures in a language. BERT is able to capture such structures, achieving a deeper query understanding than flat bag-of-words. Table~\ref{tab:query-types} also shows the limitations of BERT. It is unable to leverage evidence from negative logic conditions in narratives; removing negative conditions does not hurt performance. 


\begin{table}[tb]
 \centering
\caption{Accuracy on different types of Robust04 queries. Percentages show relative gain/loss over title queries.}\label{tab:query-types}
\def\arraystretch{0.9}
\vspace{-0.15in}
\setlength\tabcolsep{3pt}

\begin{tabular}{l|c|cr|cr|cr}
\hline \hline
\multirow{2}{*}{Query} & Avg & \multicolumn{6}{c}{nDCG@20} \\ 
  & Len & \multicolumn{2}{c}{\texttt{SDM}}   & \multicolumn{2}{c}{\texttt{Coor-Ascent}}  & \multicolumn{2}{c}{\texttt{BERT-MaxP}} \\ \hline
 



Title                   & 3   & 0.427   & -- & 0.427 & -- & 0.469    & -- \\    \hline
Desc        & 14  & 0.404   & -5\%   & 0.422  & -1\% & 0.529    & +13\% \\ 

Desc, keywords   & 7  & 0.427   & -0\%  & 0.441  & +5\%& 0.503     & +7\% \\  \hline

Narr          & 40   &  0.278  &  -35\%   & 0.424 & -1\% & 0.487  & +4\%  \\
Narr, keywords     &  18 &  0.332  & -22\%   & 0.439  & +3\%&    0.471 &  +0\%  \\ 
Narr, positive          & 31   & 0.272   &  -36\%    & 0.432 & +1\% & 0.489   & +4\%  \\

\hline \hline
\end{tabular}
\end{table}

%% file: sections/exp3-enhance.tex
\subsection{Understanding the Search Task}

Corpus-trained text representations do not always align with the search task~\cite{K-NRM}. Search specific knowledge is necessary, but requires labeled data to train.  ~\citet{dai2018convolutional} show promising results on learning search patterns from related domains for a low-resource search task. The last section investigates if BERT's language modeling knowledge can be stacked with additional search knowledge to build a better ranker, and if the search knowledge can be learned in a domain-adaptation manner to alleviate cold-start problems. 


\begin{table}[tb]
\caption{Accuracy of Bing augmented BERT on ClueWeb09-B. $\dagger$: statistically significant improvements over \texttt{Coor-Ascent}.}\label{tab:enhance}
\def\arraystretch{0.85}
\vspace{-0.15in}

\begin{tabular}{l|cc|ll}
\hline \hline
    \multirow{2}{*}{Model}  & \multicolumn{2}{c|}{Knowledge} & \multicolumn{2}{c}{nDCG@20} \\
              & Text &  Search         & {Title} & {Desc}\\ \hline
\texttt{Coor-Ascent}              & Low & Low   & 0.295     & 0.251    \\ \hline
\texttt{BERT-FirstP}  & High & Low & 0.286     & 0.272$^{\dagger}$    \\ 

\texttt{Conv-KNRM+Bing}    &  Low & High   & 0.314 $^{\dagger}$     &  0.275 $^{\dagger}$    \\ 
\texttt{BERT-FirstP+Bing}  & High  & High  & \textbf{0.333}$^{\dagger}$      & \textbf{0.300}$^{\dagger}$   \\ 
 \hline \hline

\end{tabular}
\end{table}


We train BERT on a sample of Bing search log with 0.1M queries and fine-tune it on ClueWeb09-B. Results are shown in Table~\ref{tab:enhance}. \texttt{BERT-FirstP} is the best in-domain BERT model on ClueWeb09-B (Table~\ref{tab:overall}). Its pre-trained language model encodes general word associations like (`Honda', `car'), but lacks search-specifc knowledge like  (`Honda', `special offer').
\texttt{Conv-KNRM+Bing} was the previous state-of-the-art domain adapted neural IR model~\cite{dai2018convolutional}.  It is trained on millions of query-document pairs, but does not explicitly model general language patterns. 
\texttt{BERT-FirstP+Bing}  achieves the best performance, confirming that text retrieval requires understanding both the text content and the search task. Simple domain adaptation of BERT leads to a pre-trained model with both types of knowledge that can improve related search tasks where labeled data are limited. 

%% file: sections/conclusion.tex
\section{Conclusion}
Text understanding is a long-desired feature for text retrieval.  Contextual neural language models open new possibilities for understanding word context and modeling language structures. This paper studies the effect of a recently-proposed deep neural language model, BERT, on ad-hoc document retrieval tasks.

Adapting and fine-tuning BERT achieves high accuracy on two different search tasks, showing the effectiveness of BERT's language modeling for IR. 
The contextualized model brings large improvements to natural language queries. 
The corpus-trained language model can be complemented with search knowledge through simple domain adaptation, leading to a strong ranker that models both  meanings of text and relevance in search.

People have been trained to use keyword queries because bag-of-words retrieval models cannot effectively extract key information from natural language. 
We found that queries written in natural language actually enable \textit{better} search results when the system can model language structures. Our findings encourage additional research on search systems with natural language interfaces.  
